\documentstyle[pre,aps,twocolumn,psfig,english]{revtex}
\newcommand{\fe}{f _{\rm ext}}
\newcommand{\fri}{\xi^{-1}}
\newcommand{\frii}{\xi}
\newcommand{\mup}{\mu_P}
\newcommand{\mua}{\mu_{A}}
\newcommand{\DW}{\Delta W}

\newcommand{\trans}{\omega}
\newcommand{\fp}{f_+}
\newcommand{\fm}{f_-}
\begin{document}
\title{Energy Transduction of Isothermal Ratchets: Generic Aspects
and Specific Examples Close to and Far from Equilibrium}
\author{Andrea Parmeggiani, Frank J\"ulicher, Armand Ajdari$^*$
 and Jacques Prost\\}
\date{\today}
\address{Institut Curie, Physico-Chimie Curie, UMR CNRS/IC 168, 
26 rue d'Ulm, 75248 Paris Cedex 05, France}
\address{$^*$Laboratoire de Physico Chimie Th\'eorique, Esa CNRS 7083, ESPCI,
 10 rue Vauquelin, 75231 Paris Cedex 05, France \bigskip\bigskip\\ 
\parbox{14cm}{\rm  
We study the energetics of isothermal ratchets which are driven by a
chemical reaction between two states and operate in contact with a
single heat bath of constant temperature.  We discuss generic aspects
of energy transduction such as Onsager relations in the linear
response regime as well as the efficiency and dissipation close to and
far from equilibrium.  In the linear response regime where the system
operates reversibly the efficiency is in general nonzero.  Studying
the properties for specific examples of energy landscapes and
transitions, we observe in the linear response regime that the
efficiency can have a maximum as a function of temperature.  Far from
equilibrium in the fully irreversible regime, we find a maximum of the
efficiency with values larger than in the linear regime for an optimal
choice of the chemical driving force. We show that corresponding
efficiencies can be of the order of $50\%$. A simple analytic argument
allows us to estimate the efficiency in this irreversible regime for
small external forces.
\smallskip\\
PACS Numbers: 87.10.+e, 05.40.-a}}
\maketitle
\narrowtext
\section{Introduction}
Biological systems provide an important motivation to study the
physics of active processes which on a molecular scale are able to
transduce chemical energy into mechanical work and motion.  Important
examples are motor proteins and enzymes which move actively along DNA
\cite{albe94}. The properties of such systems differ
in several respects from macroscopic machines and heat engines: (i)
active phenomena occur on a molecular scale in a very viscous
environment with overdamped dynamics, motion is thus stochastic and
obeys only on average the first and second laws of thermodynamics;
(ii) these systems are isothermal and operate strictly at constant
temperature as they are in intimate contact with a thermal bath. 
In recent years, a number of theoretical approaches to describe
this class of systems have been developed 
\cite{ajda92A,ajda92B,magn93,astu94,pros94,astu97,juli97B}.

In order to discuss the energy transduction of such systems, the
concepts which have been developed for macroscopic motors have to be
applied with some care.  Recently, there has been a growing interest
in the energetics of Brownian motors 
\cite{feyn66,magn94,juli95,seki97,shib97,soko97,juli98,mats98,dere98,parr98,hond98,kame98,seki98}. It is the aim of
this article to discuss generic aspects of energy transduction of
Brownian motors driven by a chemical reaction and to provide several
specific examples which reveal new and interesting properties.

The two-state models which we use \cite{juli97B} represent a useful
paradigm for the description of energy transduction of isothermal
motors in the overdamped regime.  They are motivated by cytoskeletal
motor proteins which move along polar and periodic filaments. Coupling
a two state model to a chemical reaction, which induces transitions
between the two states of the motor, leads to motion and
force-generation if the chemical potential difference $\Delta \mu$
between the fuel and its reaction products is nonzero and if the
system has a polar symmetry. Assuming that the chemical reservoirs
coupled to a single motor are macroscopic in size, this chemical
potential difference can be defined even under out-of equilibrium
conditions since in this limit the reaction driving the microscopic
motor affects the reservoir only weakly. Using $\Delta
\mu$ as the relevant control parameter, the consumed chemical free
energy by the active process is well defined. This leads to a simple
definition of efficiency $\eta$ as the ratio of the mechanical work
performed and the consumed chemical free energy.

We find three important results:
\begin{itemize}
\item The efficiency calculated for these
models can be maximized far from equilibrium.
\item  Close to thermal equilibrium there exists a linear response 
regime which is important because of its universal features. We 
demonstrate that the dependence of the efficiency in this regime on 
temperature is strongly model dependent and can be non-monotonous 
in which case thermal fluctuations are essential for an efficient 
energy transduction.
\item The efficiencies vanish at stalling conditions (
zero average velocity) except in a singular limit where they reach 
the ideal value $\eta=1$.
\end{itemize}

The outline of our paper is as follows.  In section II, we discuss
generic aspects which are completely independent of the model
chosen. We define the efficiency and identify the generalized currents
and forces which allow us to write a linear response theory. We
discuss the generic features of efficiency in this regime, in
particular the maximal efficiency under reversible conditions and the
efficiency at stalling conditions. In Section III we choose an
explicit realization of the transport equations where the motor is
described as a two-state model which is coupled to a chemical
reaction and we identify the energy fluxes in the system. Section IV 
discusses the
energy transduction properties for specifically chosen examples.  We
show that efficiency is typically optimized in the irreversible regime
and give examples for the temperature dependence of $\eta$ when the
system operates in the linear response regime. In our concluding
remarks, we relate our results to biological motors and discuss
alternative definitions of efficiency which have been used in the
literature.

\section{Isothermal ratchets: Generic Aspects}

\subsection{Force, velocity and efficiency}

Motivated by linear biological motor proteins which move along a
linear filament, we will consider chemically driven systems which can
induce motion along a one-dimensional track. The energy source is the
difference of the chemical potentials $\Delta \mu$ of fuel and
products.  Being motivated by biological motors, we use the hydrolysis
ATP $\rightleftharpoons$ ADP + P as example \cite{albe94}. We define
\begin{equation}
\Delta\mu=\mua-\mup
\end{equation} 
where $\mua$ and $\mup$ are the chemical potentials of ATP and ADP+P,
respectively. In order to perform useful mechanical work, the system
has to move against an external force $\fe$ applied parallel to the
track.  In addition to the two generalized forces $\Delta \mu$ and
$\fe$ acting on the system, we can define two generalized velocities:
(i) the average velocity of motion $v$ of the motor along the track;
and (ii) the chemical reaction rate $r$ defining the average number of
ATP molecules consumed per unit time.  The motor can thus be
characterized by the equations of state
\begin{eqnarray}
v & = & v(\fe,\Delta\mu) \\
r & = & r(\fe,\Delta\mu) \label{eq:eqst}
\end{eqnarray}
which describe the velocities of the system as a function of the
generalized forces \cite{footnote1}.  The mechanical work performed
per unit time against the external force is given by
\begin{equation}
\dot W = \fe v \quad .
\end{equation}
The amount of chemical energy consumed per unit time is
\begin{equation}
\dot Q = r \Delta\mu \quad .
\end{equation}
For a system which performs mechanical work, i.e. $\fe v<0$, we can
define the (mechanical) energy transduction efficiency as \cite{hill74}
\begin{equation}
\eta = -\frac{\fe v}{r \Delta\mu} \label{eq:eff}
\end{equation}
Because of energy conservation,
the amount of energy dissipated per unit time therefore
reads:
\begin{equation}
\Pi \equiv \fe v + r \Delta \mu \quad .\label{eq:diss}
\end{equation}
From the second law of thermodynamics it follows that $\Pi$ must
always be positive.

\subsection{Linear response theory}

Close to thermal equilibrium, i.e. for small forces $\fe \ll T/l$ and
$\Delta\mu\ll T$, where $l$ is a typical length scale of the motor
and $T$ is the temperature measured in units of $k_B $, we
can expand Eq. (\ref{eq:eqst}) to linear order:
\begin{eqnarray}
v & = & \lambda_{11} \fe + \lambda_{12} \Delta\mu \nonumber\\
r & = & \lambda_{21} \fe + \lambda_{22} \Delta\mu \quad .\label{eq:ons}
\end{eqnarray}
The matrix $\lambda_{ij}$ of linear response coefficients has the
following physical meaning: $\lambda_{11}$ is a mobility giving the
response of the velocity to the applied force. $\lambda_{22}$ plays a
similar role for fuel consumption. It describes the 'chemical
admittance' or the response of the chemical reaction rate $r$ to the
chemical force $\Delta \mu$. The coefficients $\lambda_{12}$ and
$\lambda_{21}$ are mechano-chemical coupling coefficients which are
responsible for energy transduction.

Looking at the symmetry of the problem, we find that $v$ and $\fe$
transform like vectors for $x\rightarrow -x$ while $r$ and $\Delta
\mu$ are scalars which do not change under inversions.  As a
consequence, the coefficients $\lambda_{11}$ and $\lambda_{22}$
transform as scalars while $\lambda_{12}$ and $\lambda_{21}$ are
vector coefficients.  The latter can be nonzero only if the system has
a polar symmetry.  Thus, the polarity of the system (polar filaments)
is essential for motion to exist.

Calculating the dissipation rate $\Pi$ in the linear regime, we find
that $\Pi$ is positive exactly if the diagonal
elements are positive, $\lambda_{ii}>0$ and if the determinant is
positive
\begin{equation}
\lambda_{11}\lambda_{22}-\lambda_{12}\lambda_{21}
>0 \quad .\label{eq:detlam}
\end{equation}
On general grounds, we expect a symmetry relation between the
Onsager coefficients if microscopic
reversibility is obeyed:
\begin{equation}
\lambda_{12}=\lambda_{21} \quad  \label{eq:ons_rel} \quad .
\end{equation}
This is a general result of non-equilibrium thermodynamics.  

\subsection{Modes of operation}
Different modes of operation of the motor can be distinguished by
looking at the input and output of energy of the system. The
dissipation rate $\Pi$ corresponds to the total flux of energy to the
thermal bath at temperature $T$. Passive regimes of the motor are
those cases where both $r \Delta \mu$ and $\fe v$ are positive: Work
performed on the system is dissipated and lost.

\begin{figure}
\centerline{\psfig{file=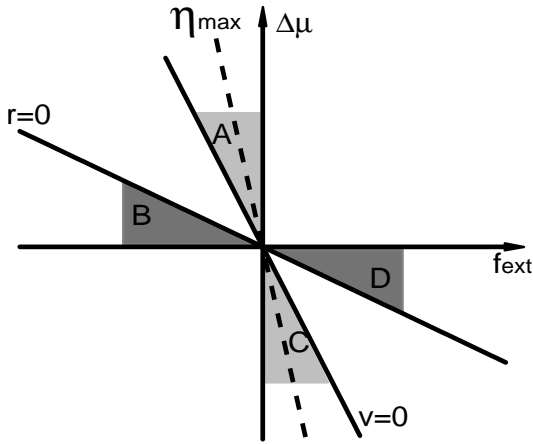,width=7.0 cm}}
\vspace{0.5cm}
\caption{Operation diagram for an isothermal motor in the linear response
regime as a function of external force $\fe$ and chemical potential
difference $\Delta\mu$. General case with four different regimes
A-D, separated by lines $v=0$ and $r=0$ where the velocity and the
fuel consumption vanish, respectively. The maximal efficiency occurs
along a line $\eta_{\rm max}$.}
\label{f:lin_regimes}
\end{figure}

More interesting are the active regimes where the motor transforms
chemical energy into mechanical work or vice versa while dissipating
only a part of the energy input. Four such active regimes exist, see
Fig. \ref{f:lin_regimes}:
\begin{itemize}
\item[A:] $r \Delta\mu >0$, $\fe v<0$, The motor uses the chemical
  energy of the ATP in excess as input and performs mechanical work
  moving with $v>0$ against a negative force $\fe<0$.
\item[B:] $r \Delta\mu <0$, $\fe v>0$,
The motor produces ATP, although already in excess, from mechanical
input due to a negative force $\fe<0$ inducing a negative velocity
$v<0$.
\item[C:] $r \Delta\mu >0$, $\fe v<0$, 
The motor uses ADP in excess to perform mechanical work.
\item[D:] $r \Delta\mu <0$, $\fe v>0$,
The motor produces ADP already in excess from mechanical work.
\end{itemize}

The different regimes are separated by the lines $\fe =0$,
$\Delta\mu=0$, $v=0$ and $r=0$.  For regimes A and C, where the motor
performs mechanical work the mechanical efficiency is the one defined
in Eq. (\ref{eq:eff}): $\eta=-\fe v/r\Delta\mu$.  Similarly, in
regimes B and D, where the system performs chemical work, the chemical
efficiency $\eta_c\equiv -r\Delta\mu/\fe v$ is more useful.

Within the linear response
regime, the efficiency can be calculated using the Onsager coefficients
\begin{equation}
\eta = -\frac{\lambda_{11} a^2+ \lambda_{12} a}
{\lambda_{21}a+\lambda_{22}} \quad ,
\label{eq:etaml}
\end{equation}
where $a=\fe/\Delta\mu$. If we choose a constant $\Delta\mu>0$, the
efficiency vanishes for $\fe=0$ (no work is performed).  $\eta$
becomes positive for $\fe<0$ (note the minus sign which indicates that
the force is applied in the direction opposing movement), reaches a
maximum for a certain value of the force and becomes zero again at the
stall force for which $v=0$.  According to Eq. (\ref{eq:etaml}), the
efficiency is constant along straight lines $\fe=a \Delta\mu$ which
correspond to constant $a$. Thus, at the origin of the
$(\fe,\Delta\mu)$-plane which corresponds to thermal equilibrium and
reversible, quasistatic operation, the efficiency $\eta$ has a
singularity and is multi-valued.

Maximal efficiency occurs for a certain value $a$ for
which $\partial \eta/\partial a=0$. It is given by
\cite{kede65,juli97B}
\begin{equation}
\eta_{\rm max}=(1-\sqrt{1-\Lambda})^2/\Lambda \quad .\label{eq:lin_eta}
\end{equation}
Here, $\Lambda \equiv \lambda_{12}^2/(\lambda_{11}\lambda_{22})$.  It
varies between $\eta_{\rm max}=0$ for $\lambda_{12}=0$ and $\eta_{\rm
max}=1$ if $\lambda_{12}^2=\lambda_{11}\lambda_{22}$.  Larger values
$\Lambda>1$ violate thermodynamics according to Eq. (\ref{eq:detlam})
and the Onsager relation (see Eq. (\ref{eq:ons_rel})).

These arguments demonstrate that the efficiency vanishes under
stalling conditions $v=0$. This is an important difference from Carnot
engine for which the efficiency is optimized under quasistatic
conditions without net motion. It results from the fact that the
energy transduction driven by a chemical reaction considered here will
in general still have a nonzero consumption rate $r$ even when motion
stops, or in other words $v=0$ and $r=0$ do not occur for the same
conditions.

There is however one limiting case where this is no longer true: If
$\Lambda\rightarrow 1$, the two lines $r=0$ and $v=0$ in the
$(\fe,\Delta\mu)$-plane tend towards each other.
In this limit the chemical reaction and motion are strictly coupled
(i.e. one can't occur without the other) and the efficiency reaches
the maximum $\eta=1$. This situation is an idealized case which
applies to good approximation to polymerization forces and motion
generated by polymerization processes as in the case of RNA polymerase
\cite{dogt97,juli98}

\section{Two state model}

\subsection{Transport equations}

We study energy transduction and efficiencies of isothermal motors
using simple two-state models. The motor is characterized by its
position $x$ along a one dimensional coordinate describing the polar
and periodic track. We assume that the motor exists in two different
conformations or states $\sigma=1$,$2$.  The interaction between motor
and track depends on $\sigma$ and is described by potentials
$W_\sigma(x)$ with polar symmetry which are periodic with period $l$.

The role of the chemical reaction is to trigger transitions between
the two states. We introduce the position dependent rate constants
$\omega_{1}(x)$ and $\omega_2(x)$ which characterize the probability
per unit time for the transitions $1\rightarrow 2$ and $2\rightarrow
1$ at position $x$, respectively. The probability densities $P_1(x,t)$
and $P_2(x,t)$ for the system to be at time $t$ at position $x$ in one
of the two states obey the Fokker-Planck Equations \cite{pros94}
\begin{eqnarray}
\partial_t P_1 + \partial_x J_1& =& -
\omega_1(x) P_1 + \omega_2(x) P_2 \nonumber\\
\partial_t P_2 + \partial_x J_2& =& 
\omega_1(x) P_1 - \omega_2(x) P_2 \quad  \label{eq:Psp}
\end{eqnarray}
The particle currents are given by
\begin{equation}
J_\sigma \equiv \fri [-T\partial_x P_\sigma
- P_\sigma \partial_x W_\sigma + P_\sigma \fe ] \quad ,\label{eq:J}
\end{equation}
where $\fri$ is an effective mobility, the temperature $T$ is measured
in units of $k_B$ and $\fe$ is the external force introduced above.

For given rates $\omega_\sigma$ the system relaxes to a steady state
with $\partial_t P_\sigma=0$. The normalized distributions which
satisfy periodic boundary conditions ($\int_0^l dx\; (P_1+P_2)=1$,
$P_\sigma(0)=P_\sigma (l)$ and $\partial_x P_\sigma(0)=\partial_x
P_\sigma (l)$) in the steady state allow us to calculate the average
velocity
\begin{equation}
v=\int_0^ldx\;(J_1+J_2) \quad .\label{eq:v}
\end{equation}

\subsection{Coupling to a chemical reaction}

We now consider the situation where the transitions between states $1$
and $2$ occur as a result of a chemical reaction scheme which we model
separately.  In order to be general and to capture different
situations, we consider the following scheme:
\begin{eqnarray}
ATP + M_1 &  \begin{array}{c c c}
 &\alpha_1 & \\
 & \rightleftharpoons & \\
 & \alpha_2 & 
\end{array}
& M_2 + ADP + P \\
ADP+P + M_1 &  \begin{array}{c c c}
 &\gamma_1 & \\
 & \rightleftharpoons & \\
 & \gamma_2 & 
\end{array}
& M_2 + ATP \\
M_1  & \begin{array}{c c c}
 &\beta_1 & \\
 & \rightleftharpoons & \\
 & \beta_2 & 
\end{array}
&  M_2  \quad ,
\end{eqnarray}
where $\alpha_i$, $\gamma_i$ and $\beta_i$ denote the forward and
backward rates, respectively. The reaction pathway $\alpha$ involves
ATP hydrolysis with chemical free energy gain $\Delta\mu$ when
changing from state $1$ to state $2$, while pathway $\gamma$ involves
hydrolysis in the opposite direction. The transitions $\beta$ are do
not involve a chemical potential difference.  Chemical kinetics requires
\begin{eqnarray}
\frac{\alpha_1}{\alpha_2}& =& e^{(W_1-W_2+\Delta\mu)/ T} \\
\frac{\gamma_1}{\gamma_2}& =& e^{(W_1-W_2-\Delta\mu)/ T} \\
\frac{\beta_1}{\beta_2}& =& e^{(W_1-W_2)/ T}\quad .
\end{eqnarray}
The transition rates can therefore be written as 
\begin{eqnarray}
\omega_1 & = &\! \alpha_2 e^{(W_1-W_2+\Delta\mu)/ T}\!+\!\gamma_2 e^{(W_1-W_2-\Delta\mu)/ T}\!+\!\beta_2 e^{(W_1-W_2)/ T} \nonumber \\
\omega_2 & = &\! \alpha_2\! +\! \gamma_2\!+\!\beta_2 \quad , \label{eq:om1om2}
\end{eqnarray}
where unknown ($l$-periodic) functions $\alpha_2(x)$, $\gamma_2(x)$
and $\beta_2(x)$ define the conformation dependence of transitions
rates \cite{footnote2}.  With these expressions, the net steady state
ATP consumption rate is given by
\begin{equation}
 r = \int_0^l dx 
\left [ (\alpha_1(x)-\gamma_1(x)) P_1(x)- 
(\alpha_2(x)-\gamma_2(x)) P_2(x) \right ] .\label{eq:r}
\end{equation}

\subsection{Detailed balance}

If $\Delta\mu=0$, the chemical reaction is in equilibrium and the
transition rates are just thermal fluctuations and obey the relation
of detailed balance $\omega_1/\omega_2 =
\exp((W_1-W_2)/T)$. Breaking of detailed balance for $\Delta\mu\neq 0$
is a requirement for spontaneous motion and force generation to be
possible. In order to quantify the departure from thermal equilibrium
and the extend to which detailed balance is broken, we define the
quantity
\begin{equation}
\Omega(x) = \omega_1(x)
- \omega_2(x) \exp\left(-\frac{\DW(x)}{T}\right) \quad ,
\end{equation}
with $\DW(x)=W_2(x)-W_1(x)$. Detailed balance is obeyed only 
if $\Omega(x)=0$ for all $x$.
Using the transition rates as given by Eq. 
(\ref{eq:om1om2}), we find
\begin{equation}
\Omega(x) =  e^{-\Delta W/T} \left [\alpha_2(x)\left(e^{\Delta\mu/T}-1\right)
+ \gamma_2(x)\left(e^{-\Delta\mu/T}-1\right)\right ] .
\end{equation}

If $\Delta\mu\neq 0$, we distinguish two interesting limits: for small
$\Delta\mu/T\ll 1$
\begin{equation}
\Omega(x)\simeq (\alpha_2(x)-\gamma_2(x)) 
e^{-\Delta W/T} \frac{\Delta\mu}{T} \quad ,
\label{eq:Odmu}
\end{equation}
indicating that $\Omega$ is proportional to $\Delta \mu$.
If $\Delta\mu/T$ is large compared to one,
$\Omega$ depends only on the ratio $k=[ATP]/[ADP][P]$:
\begin{equation}
\Omega(x)\simeq  (\alpha_2(x)e^{\Delta\mu^0/T}  k
+ \gamma_2(x)e^{-\Delta\mu^0/T}k^{-1}) \quad ,
\end{equation}
where $ \Delta \mu^0 = \mu^0_{ATP} -\mu^0_{ADP} -\mu^0_{P}$.  Here, we
used the relation $\mu_i=\mu_i^0+T \ln [i]$ where $[i]$ is the
concentration of species $i$, and $\mu_i^0$ the so called standard
chemical potential.

\subsection{Energy conservation and dissipation}
\label{s:diss}
The first law of thermodynamics requires that the energy flow
through the system is conserved as described by Eq. (\ref{eq:diss}).
This energy conservation can be derived from the transport
equations. This leads to expressions for the local density
of energy dissipation which gives interesting insights in how
energy transduction is occurring.

We distinguish two types of dissipation rates: (i) the dissipation
rates $\Pi_\sigma$, with $\sigma=1,2$ corresponding to sliding within
the potential profiles and (ii) the dissipation rates $\Pi_\mu$, with
$\mu=\alpha,\beta,\gamma$ corresponding to transitions between the two
states. In addition to the total dissipation rates $\Pi_\sigma$ and
$\Pi_\mu$, we introduce local dissipation densities $\Theta_\sigma(x)$
and $\Theta_\mu(x)$ with $\Pi=\int_0^l dx
\Theta(x)$.

For a particle sliding in the potential $W_\sigma(x)$ with a steady state
distribution $P_\sigma(x)$ \cite{lebo55,degr62} 
\begin{equation}
\Theta_\sigma= - J_\sigma \partial_x H_\sigma \label{eq:dissdens}
\end{equation}
and
\begin{equation}
\Pi_{\sigma} = - \int_0^l dx
\; J_\sigma(x) \partial_x H_\sigma(x) \quad , \label{eq:dissloc}
\end{equation}
where
\begin{equation}
H_\sigma(x) \equiv  W_\sigma(x) -\fe x + T \ln (P_\sigma(x)) 
\end{equation}
is an enthalpy whose gradient induces the
Fokker-Planck current $J_\sigma$:
\begin{equation}
 J_\sigma= - \fri P_\sigma\partial_x H_\sigma \quad . 
\end{equation}
Therefore, $\Pi_\sigma$ is positive definite as expected for a
dissipation rate.  Similarly, the dissipation densities corresponding
to chemical transitions are given by
\begin{eqnarray}
\Theta_\alpha &=& (\alpha_1 P_1 -\alpha_2 P_2)
(H_1-H_2+\Delta\mu) \nonumber \\
\Theta_\gamma &=& (\gamma_1 P_1 -\gamma_2 P_2)
(H_1-H_2-\Delta\mu) \nonumber \\
\Theta_\beta &=& (\beta_1 P_1 -\beta_2 P_2)
(H_1-H_2) \quad ,
\end{eqnarray}
and 
\begin{equation}
\Pi_{\mu} = \int_0^l dx\; \Theta_\mu(x) \quad ,
\label{eq:diss_th}
\end{equation}
for $\mu=\alpha,\beta,\gamma$.  Again, $\Pi_{\alpha}$, $\Pi_{\beta}$
and $\Pi_{\gamma}$ are positive definite as required.  For a steady
state with periodic boundary conditions, we can partially integrate
Eq. (\ref{eq:dissloc}) and find together with Eq. (\ref{eq:Psp})
\begin{equation}
\begin{array}{lll}
\Pi_1 + \Pi_2  & = & \int_0^l dx\; (H_1 \partial_x J_1 + 
H_2 \partial_x J_2) +\fe v \\
 & = & \int_0^l dx\;
(H_1-H_2) (\omega_1 P_1 - \omega_2 P_2) +\fe v\quad .
\end{array}
\end{equation}
Using Eqns. (\ref{eq:J}),(\ref{eq:om1om2}) and (\ref{eq:r}), we find
that, the total dissipation rate 
\begin{equation}
\Pi=\Pi_1+\Pi_2+
\Pi_\alpha+\Pi_{\beta}+\Pi_{\gamma}\label{eq:pi12abc}
\end{equation}
satisfies Eq. (\ref{eq:diss}) and energy
conservation is obeyed.

For small $\Delta\mu$ and small $\fe$, the two state model has a
linear response regime which obeys the general properties
required by thermodynamics. In particular it can be demonstrated
that the model satisfies the symmetry relation of
Eq. (\ref{eq:ons_rel}) as we describe in appendix \ref{s:ons_rel}.
\begin{figure}
\centerline{\psfig{file=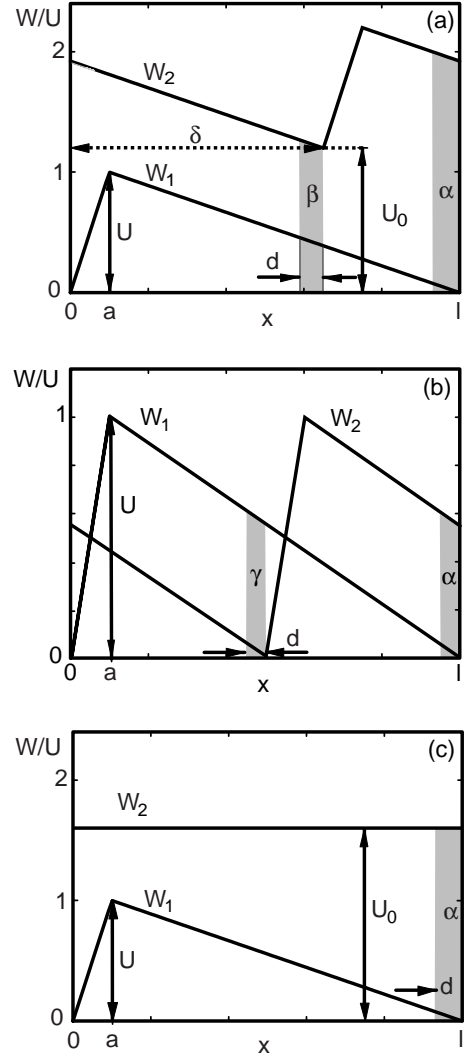,width=6.0cm}}
\vspace{0.5cm}
\caption{Three choices of potentials $W_1$ and $W_2$ with period $l$
and transition regions indicated in grey.  The position $a$ of the
maximum of $W_1$ characterizes the potential asymmetry, $U$ denotes
the potential amplitudes.  (a) System A with potentials shifted by a
distance $\delta$ and offset $U_0$. Active transitions $\alpha$ and
thermal transitions $\beta$ are localized within regions of size $d$
near the potential minima. (b) System B with symmetric states. The
potentials are shifted by a distance of $l/2$, active transitions
$\alpha$ and $\gamma$ are chosen such that the system is symmetric
with respect to an exchange of the two states.  (c) System C with a
flat potential $W_2$, localized active transitions $\alpha$ and
non-localized thermal transitions $\beta$.}
\label{f:pot}
\end{figure}

\section{Efficiencies close to and far from equilibrium}

\subsection{Specific examples}

We have introduced a general framework which allows us to study a
large variety of systems which differ in their potential shapes and in
the transition rates $\alpha_2$, $\gamma_2$ and $\beta_2$. We now
discuss three particular examples which we have chosen as prototypes
to illustrate the physics of energy transduction.

{\bf System A} is a system with two periodic potentials of equal
amplitude $U$ which are piecewise linear and which are shifted with
respect to each other by a displacement $\delta$ as shown
schematically in Fig. \ref{f:pot} (a).  
Furthermore, they differ by a constant value $U_0$: $W_2(x)=
W_1(x-\delta)+U_0$. The potentials are characterized by the asymmetry
parameter $a$ which denotes the position of the potential maximum of
$W_1$.  We choose a reaction scheme with chemically activated
transitions $\alpha_{1,2}$ between the low energy state $1$ and the
high energy state $2$, passive transition $\beta_{1,2}$ and
$\gamma_{1,2}=0$. The chemical cycle corresponds to subsequent
transitions $\alpha$ and $\beta$ which we choose localized within
intervals of size $d$:
\begin{equation}
\alpha_2(x) = \left\{
\begin{array}{l l}
\trans & \quad {l-d\leq x\leq l} \\
0 & \quad \hbox{\rm otherwise,}
\end{array}
\right .\label{eq:alpha2}
\end{equation}
localized near the minimum of $W_1$ and
$\beta_2(x)=\alpha_2(x-\delta)$ localized near the minimum of $W_2$,
see Fig. \ref{f:pot} (a). Here we have for simplicity introduced a 
single parameter $\trans$ which sets the typical time scale of 
transition rates.  
The transition rates $\omega_\sigma$ of system A obey
\begin{eqnarray}
\omega_1(x) =\alpha_2(x) &e^{(W_1-W_2+\Delta\mu)/T} &+ \alpha_2(x-\delta)
e^{(W_1-W_2)/T} \nonumber \\
\omega_2(x) =\alpha_2(x) & &+ \alpha_2(x-\delta) 
\quad .
\label{eq:tr_modA}  
\end{eqnarray}
System A is chosen in such a way that diffusion within the
potentials is not necessary for motion generation and each chemical
cycle generates with high probability a forward step along the
$x$-coordinate.

{\bf System B} has different symmetry and different topology of the
chemical reaction scheme as compared to system A, see Fig
\ref{f:pot} (b).  The two potentials are shifted by exactly half a
potential period $\delta=l/2$: $W_2=W_1(x-l/2)$ and $U_0=0$.  This
allows us to introduce a new symmetry: the system is invariant under a
shift $x\rightarrow x+l/2$ if at the same time the states are
exchanged: $1\rightarrow 2$. This situation is realized by choosing
transition rates $\beta_{1,2}=0$ and $\gamma_1(x)=\alpha_2(x-l/2)$
where we localize all transitions near the potential minima.  We can
therefore write for system B
\begin{eqnarray}
\omega_1(x) =&\alpha_2(x)e^{(W_1-W_2+\Delta\mu)/T} + 
\alpha_2(x-l/2)& \nonumber \\
\omega_2(x) =&\alpha_2(x) + \alpha_2(x-l/2) e^{(W_2-W_1+\Delta\mu)/T}&
 \quad , \label{eq:tr_modB}  
\end{eqnarray}
with $\alpha_2(x)$ given by Eq. (\ref{eq:alpha2}). Note, that system
B involves {\em two} active chemical steps per potential period.
However, because of its additional symmetry it is $l/2$-periodic.
Furthermore, all chemical transitions involve ATP hydrolysis, there
are no passive transitions.

{\bf System C} is shown in Fig. \ref{f:pot} (c). It is a variant of model
A with a weakly bound state $W_2(x)=U_0$ of constant energy. As for
system A we choose a reaction scheme with $\gamma_{1,2}=0$ and
localized active transitions near the minima using again definition
(\ref{eq:alpha2}). Since the potential $W_2$ is structureless, we
assume passive transitions to be non-localized with
$\beta_2(x)=\omega$.  Therefore in system C
\begin{eqnarray}
\omega_1(x) =\alpha_2(x) &e^{(W_1-W_2+\Delta\mu)/T} &+ \trans 
e^{(W_1-W_2)} \nonumber \\
\omega_2(x) =\alpha_2(x) & &+ \trans
\quad .
\label{eq:tr_modC}  
\end{eqnarray}
In this case motion generation involves a diffusive step in
state $2$ which we expect to reduce the efficiency of energy
transduction.

In order to discuss these models, we identify the relevant
dimensionless parameters: the dimensionless position $\bar
x=x/l$, reduced temperature $t=T/U$, reduced potentials
$w_\sigma=W_\sigma/U -\fe l/U$ and reduced transition rates
$\bar\omega_\sigma=\omega_\sigma/\trans$. Eqns. (\ref{eq:Psp}) and
(\ref{eq:J}) can for a steady state be written as
\begin{eqnarray}
-\partial_{\bar x}(t\partial_{\bar x} P_1+
P_1\partial_{\bar x}w_1 )&=& \chi(-\bar\omega_1 P_1 +\bar\omega_2 P_2)
\nonumber  \\
-\partial_{\bar x}(t\partial_{\bar x} P_2+
P_2\partial_{\bar x}w_2 )&=& \chi(\bar\omega_1 P_1 -\bar\omega_2 P_2) 
\quad . \label{eq:adim}
\end{eqnarray}
The dimensionless parameter
\begin{equation}
\chi\equiv \frac{\trans \frii l^2}{U} \label{eq:chi}
\end{equation}
compares two time-scales: (i) the typical chemical time $\trans^{-1}$
and (ii) the typical sliding time in the potentials $\frii l^2/U$.
For $\chi\gg 1$ transitions are fast compared to sliding while for
$\chi\ll 1$ sliding is fast. The model is fully characterized by the
dimensionless parameters $\chi$, $T/U$, $\Delta\mu/U$, $a/l$, $d/l$, $\delta/l$
and $U_0/U$.  The results discussed in the following section are
obtained by numerically solving Eq. (\ref{eq:adim}) with periodic
boundary conditions for the three different systems.

\subsection{Efficiencies close to equilibrium}

\begin{figure}
\centerline{\psfig{file=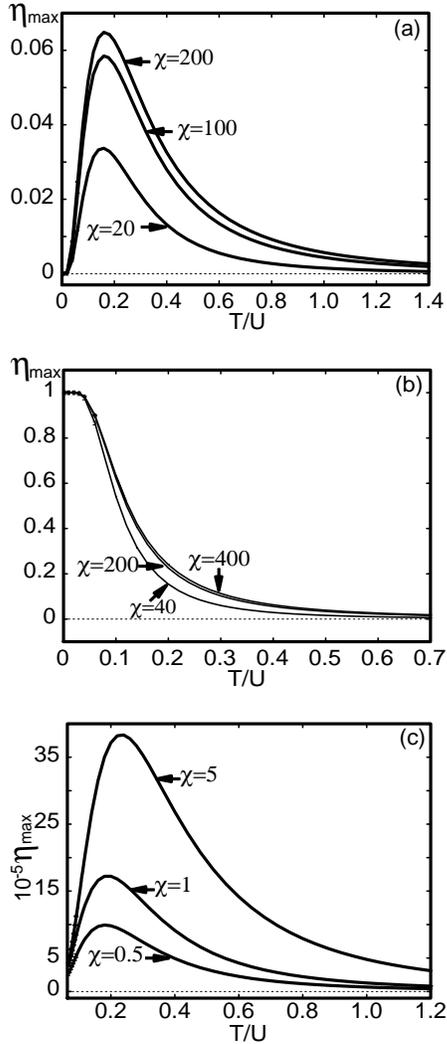,width=6.0cm}}
\vspace{0.5cm}
\caption{Maximal efficiency $\eta_{\rm max}$ in the linear response
regime as function of reduced temperature $T/U$ 
for systems A, B and C with $a/l=0.1$ as shown in Fig.
\protect\ref{f:pot}. (a) System A with $\delta/l=0.65$, 
$U_0/U=0.4$, at different $\chi$. (b) Same diagram for system B at
different $\chi$. (c) Same diagram for system C with $U_0/U=1.2$ at
different $\chi$.}
\label{f:efflin}
\end{figure}

Numerical examples for the maximal efficiency in the linear response
regime as a function of temperature are displayed in
Fig. \ref{f:efflin} for systems A, B and C and different values
of $\chi=\trans\frii l^2/U$.  
They have been obtained by first
calculating Onsager coefficients from steady state solutions for small
$\Delta\mu$ and small $\fe$ and using Eq. (\ref{eq:lin_eta}).  The
orders of magnitude of the efficiency differ for systems A, B and
C. The efficiency $\eta$ depends on $\chi$ and increases in general
with increasing $\chi$.  System B has the largest efficiency which
approaches $\eta\simeq 1$ for small $T/U$ and decreases monotonically
as a function of temperature.  For systems A and C the efficiency
has a maximum as a function of temperature and vanishes in the limit
of small $T/U$. This indicates in these cases the importance of
thermal fluctuations for energy transduction.  Note, that the limit of
small temperatures is subtle since in linear response $\Delta\mu \ll
T$ must be obeyed.  Therefore, this limit corresponds to first sending
$\Delta\mu$ to zero and $T$ afterwards. Even for small temperatures
the system thus remains in a regime where thermally activated passage
over energy barriers rests important.  While system A can have
significant efficiencies of the order of $\eta\simeq 0.06$ in the
linear response regime, the efficiency of system C which relies on
diffusive steps is small ($\eta\simeq 10^{-4}$), see
Fig. \ref{f:efflin}.
\begin{figure}
\centerline{\psfig{file=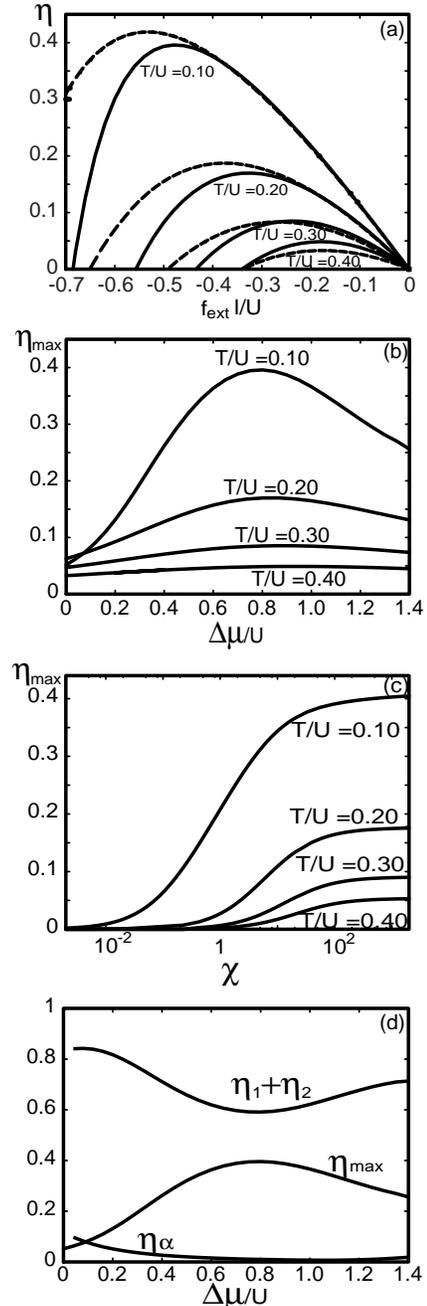,width=5.5cm}}
\vspace{0.0cm}
\caption{
Energy transduction of system (A) with $a/l=0.1$ and $U_0/U=0.4$.  (a)
Efficiency $\eta$ as a function of the external force $\fe$ for
$\Delta\mu/U=0.8$, $\chi=\xi l^2 \trans/U=200$, and different reduced
temperatures $T/U$.  Broken lines represent the approximation
discussed in the text.  (b) Maximal efficiency $\eta_{\rm max}$ as a
function of $\Delta\mu/U$ for $\chi=200$ and different temperatures.
(c) Maximal efficiency as a function of $\chi$ for $\Delta\mu/U=0.8$
and different temperatures.  (d) Relative dissipation rates as a function of
$\Delta\mu$ for the same system: shown are the fraction of energy
dissipated by potential sliding $\eta_1+\eta_2$ and the fraction
dissipated via active transitions $\eta_\alpha$. For details see
text.}
\label{f:eff_nl}
\end{figure}

\subsection{Efficiencies far from equilibrium}

We have shown that the two state model transduces chemical energy into
mechanical work in the linear response regime, however with varying
efficiencies. 
In linear response the chemical action represents a
small bias of the dominant thermal fluctuations. We are now comparing
these results with the properties of energy transduction far from
equilibrium.

{\bf System A}: Fig. \ref{f:eff_nl} (a) displays the efficiency $\eta$
as a function of the applied force for the system A as defined in
Fig. \ref{f:pot} for $\Delta\mu/T=8$ and different temperatures. 
The efficiency vanishes for $\fe=0$ as well as for the stall force for
which the velocity vanishes.
For an intermediate value of the force,
the efficiency reaches a maximum.  This value $\eta_{\rm max}$ is
displayed in Fig. \ref{f:eff_nl} (b) as a function of $\Delta
\mu$. This diagram reveals the main characteristics of energy
transduction: for small $\Delta\mu$ we find again the non-vanishing
efficiency of the linear regime.  The efficiency increases as a
function of $\Delta\mu$, reaches a maximal value and decreases for
large $\Delta\mu$ to zero. For sufficiently large values of
$\Delta\mu$ the efficiency increases for decreasing temperatures and
reaches in the example shown a value of $\eta\simeq 0.4$ for
$T/U\simeq 0.1$.  
The results obtained for different temperatures
intersect for small $\Delta\mu$ which corresponds to the observation
discussed above that the efficiency in the linear regime displays a
maximum as a function of temperature.  Fig. \ref{f:eff_nl}(c) shows
the behavior of $\eta_{max}$ for fixed $\Delta\mu/U$ as a function of
$\chi=\xi l^2\trans/U$ over a range of $6$ decades. 
The efficiency
increases monotonically with increasing $\chi$ from zero to a plateau
value. As an important result we find that the largest values of the
efficiency for the relevant energy scale $T/U\simeq 0.1$ are of the
order of $\eta\simeq 0.5$ and occur for $\Delta\mu\sim U$ comparable
to the energy difference between the two states at the
transition and thus far from the linear regime.

The dissipation rate $\Pi$ can, according to Eq. \ref{eq:pi12abc}, be
divided into separate contributions of potential sliding $\Pi_\sigma$
and chemical transitions $\Pi_\mu$, with $\mu=\alpha,\beta,\gamma$.
It is useful to define relative dissipation rates $\eta_\sigma =
\Pi_\sigma/r\Delta\mu$ and $\eta_\alpha=\Pi_\alpha/r\Delta\mu$ which
are analog to the efficiency and describe the fraction of dissipated
energy relative to the consumed chemical work.  Note that
$\eta+\eta_1+\eta_2+\eta_\alpha+\eta_\beta+\eta_\gamma=1$ follows from
energy conservation.  Fig. \ref{f:eff_nl} (d) shows the dominant
relative dissipation rates together with the efficiency $\eta$.  The
dominant dissipation is $\Pi_1+\Pi_2$ resulting from potential
friction, dissipation $\Pi_\alpha$ of chemical transitions, plays a
minor role. The dissipation $\eta_\beta$ corresponding to passive
transitions is smaller than $0.01$ and can be neglected.  It is
therefore not shown. The maximum of $\eta$ corresponds to a minimum of
$\eta_1+\eta_2$.

\begin{figure}
\centerline{\psfig{file=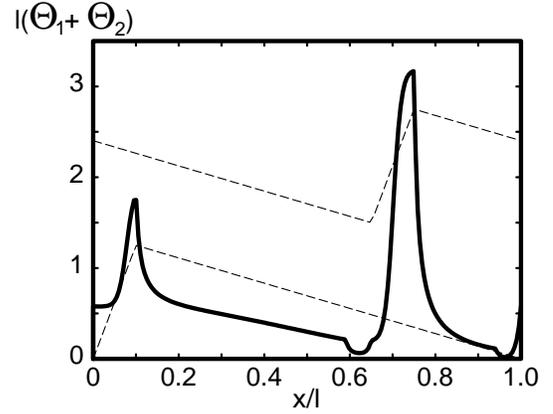,width=7.0 cm}}
\vspace{0.5cm}
\caption{Density $\Theta_1(x)+\Theta_2(x)$ as defined in 
Eq. (\protect\ref{eq:dissdens}) of the dissipation rate as a function
of position $x$ for system A with $Ul/\fe=-0.48$, $\chi=200$,
$\Delta\mu/U=0.80$.}
\label{f:diss_loc}
\end{figure}

The main energy loss results
from thermally activated backward steps. This idea can be directly
tested by calculating the local dissipation density
$\Theta_1(x)+\Theta_2(x)$ as defined in Eq. (\ref{eq:dissdens}).  This
quantity is displayed in Fig. \ref{f:diss_loc}. 
The plot reveals that maximal dissipation occurs for
$\delta-a<x<\delta$, i.e. along the steep potential slope of the
potential $W_2$.  
A second maximum of local dissipation exists for
$0<x<a$ along the steep slope of $W_1$. In contrast, minimal
dissipation occurs near the potential minima where transitions between
states take place. The steep potential slopes where the density of
energy dissipation is large indeed are accessed via thermally
activated backward hopping events.  The probability of such events
increases in the presence of an ``adverse'' external force which
limits the efficiency of the system.

\begin{figure}
\centerline{\psfig{file=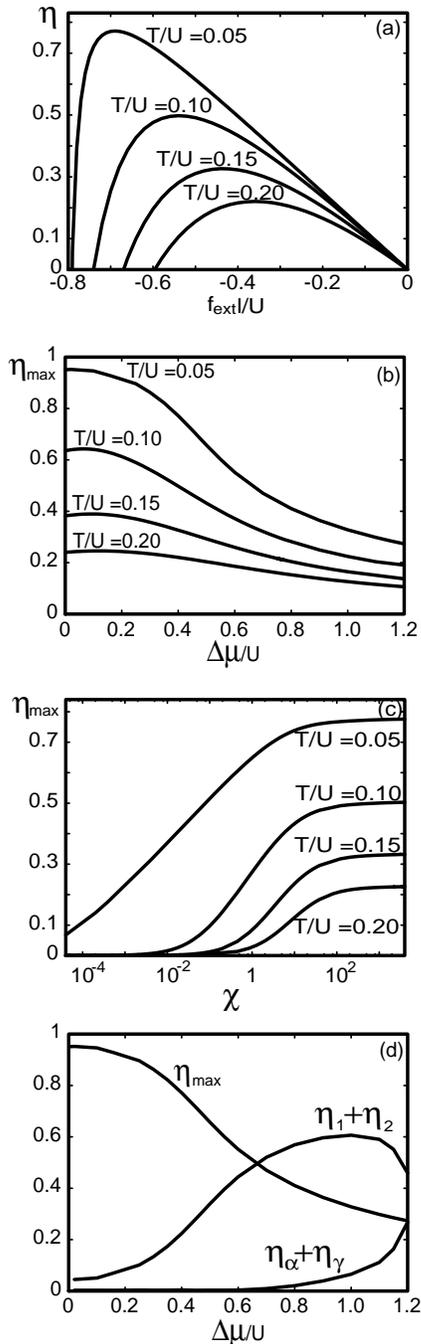,width=5.5cm}}
\vspace{0.0cm}
\caption{
Energy transduction of system B with $a/l=0.1$.  (a) Efficiency
$\eta$ as a function of the external force $\fe$ for
$\Delta\mu/U=0.4$, $\chi=\xi l^2 \trans/U=400$, and different reduced
temperatures $T/U$.  (b) Maximal efficiency $\eta_{\rm max}$ as a
function of $\Delta\mu/U$ for $\chi=400$ and different temperatures.
(c) Maximal efficiency as a function of $\chi$ for $\Delta\mu/U=0.4$
and different temperatures.  (d) Relative dissipation rates as a function of
$\Delta\mu$: the fraction of energy dissipated by potential sliding
$\eta_1+\eta_2$ and the fraction dissipated via active transitions
$\eta_\alpha + \eta_\gamma$.}
\label{f:eff_nl_kin}
\end{figure}

{\bf System B}: Fig. \ref{f:eff_nl_kin} displays the same information
for system B.  The diagram reveals that efficiencies are in general
larger than for system A, reaching values up to $\eta\simeq 0.7$ for
$T/U=0.05$. 
Furthermore, the maximum of the efficiency as a function
of $\Delta\mu$ is less pronounced and shifted to small values of
$\Delta\mu$ as compared to system A.  One might expect that the
dissipation due to passive transitions $\eta_\beta$ in system A which
does not exist in system B could play a role in improving the
efficiency of system B. However as discussed above $\eta_\beta$ can be
neglected and is thus not responsible for this effect.  The main
reason for the improved efficiency of system B is the fact that the
effective energy barrier for thermally activated passage over the
potential maxima is larger in system B as compared to system A.
Therefore, fluctuations leading to ``backward steps'' in the opposite
direction of average motion which completely dissipate a consumed ATP
molecule are less likely.  Each active chemical transition is thus
transduced into work with high probability.  Fig. \ref{f:eff_nl_kin}
(c) shows qualitatively the same behavior of the efficiency as a
function of $\chi$ for system B as compared to system A.  Also as
discussed for system A, the dominant dissipation process
corresponds to sliding in the potentials, see
Fig.\ref{f:eff_nl_kin} (d). Note that the efficiency is larger than in 
system A, which correlates with the increased barrier height reducing 
the probability of backward steps.

\begin{figure}
\centerline{\psfig{file=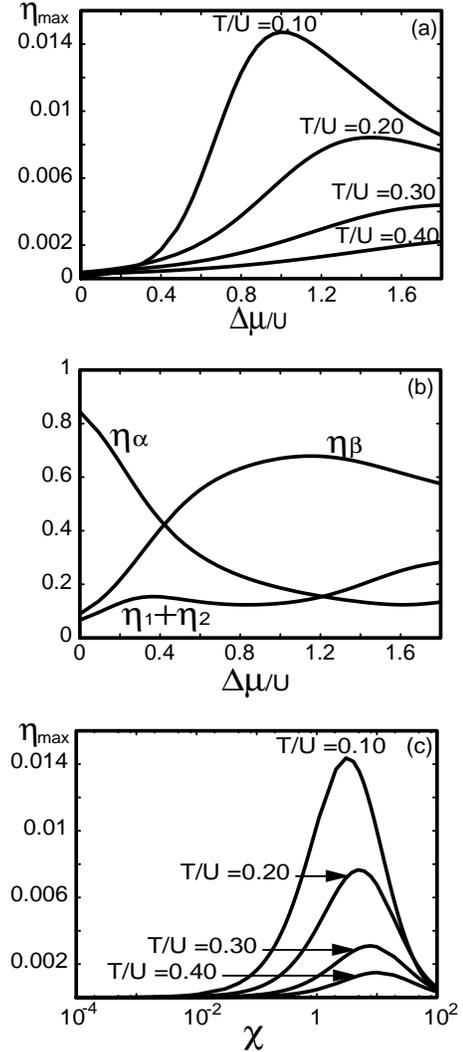,width=6.0cm}}
\vspace{0.5cm}
\caption{
Energy transduction of system C with $a/l=0.1$.  (a) Maximal
efficiency $\eta$ as a function of $\Delta\mu/U$ for $\chi=\xi l^2
\trans/U=5$, $U_0/U=1.2$ and different reduced temperatures $T/U$.
(b) Relative dissipation $\eta_1+\eta_2$ in the potentials as well
as the dissipation of transitions $\eta_\alpha$ and $\eta_\beta$
corresponding to (a).  (c) Maximal efficiency as a function of $\chi$
for $\Delta\mu/U=1.2$ and different temperatures.}
\label{f:eff_nl_flat}
\end{figure}

{\bf System C}: Energy transduction of system C which involves
diffusive steps and non-localized de-excitations. 
Maximal efficiencies
are of the order of $0.02$ and thus much smaller than those for
systems A and B, see Fig. \ref{f:eff_nl_flat}. 
As in system A the largest efficiencies occur for $\Delta\mu\gg T$ and
thus far from equilibrium. The reason for the reduced efficiency
becomes clear when studying the relative dissipation rates shown in
Fig. \ref{f:eff_nl_flat} (b): Most energy is in this case dissipated
by the passive and active transitions, potential sliding is less
important. In particular, the non-localized and passive de-excitations
dominate dissipation far from equilibrium.  
Very striking is the
behavior of the efficiency as a function of $\chi$ shown in Fig.
\ref{f:eff_nl_flat} (c): The efficiency displays a maximum for
certain values of $\chi$ but vanishes both for large and small
$\chi$. This property reflects the fact that a matching of
time scales is crucial for this system: The life-time in the
excited state should be comparable to the diffusion-time over a potential 
period:
\begin{equation}
l^2\sim T /\frii \trans \quad.
\end{equation}
Therefore, the optimal value of $\chi$ should behave as $\chi_{\rm
opt} \sim T/U$ which explains the temperature-dependence of the
maximum in Fig. \ref{f:eff_nl_flat} (c).

\subsection{Approximation for small forces}
\label{s:smallf}

\begin{figure}
\centerline{\psfig{file=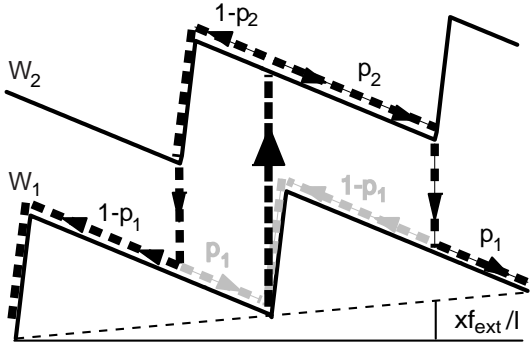,width=7.0 cm}}
\vspace{0.5cm}
\caption{Schematic diagram of events after consumption
of one fuel molecule. Forward steps occur with probability $p_+=p_1
p_2$, backward steps with probability $p_-=(1-p_2)(1-p_1)$ and neutral
steps with $p_0=p_2(1-p_1)+p_1(1-p_2)$ .}
\label{f:prob}
\end{figure}

The efficiency far from equilibrium for $\Delta\mu/T\gg 1$ but for
small forces can be understood by a simple approximation which we
discuss for system A.  In the limit of large $\Delta\mu$ and $U/T$ we
ignore spontaneous hopping events over the maxima of potential
$W_1$. 
Every ATP consumption event corresponds to a transition to the
second state from which the particle will eventually decay to the
first state.  During this process, it undergoes a forward step with
probability $p_+$, a backward step with probability $p_-$ or it will
return to the initial position with probability $p_0$. Here, we have
ignored multiple steps, see Fig. \ref{f:prob}. 
In the presence of an external force $\fe$, the efficiency can thus be
estimated as
\begin{equation}
 \eta \simeq  -\frac{\fe}{\Delta\mu} <x> \quad ,
\end{equation}
where 
\begin{equation}
<x>\simeq l(p_+ - p_-) \simeq v/r \label{eq:etaapp}
\end{equation}
is the average displacement per consumed ATP.  The probabilities
$p_\pm$ can be written as
\begin{equation}
p_+= p_2(0) p_1(\delta)  \quad , 
\quad p_-=(1-p_2(0))(1-p_1(\delta)) \quad . \label{eq:ppm}
\end{equation}
Here, we have introduced the probabilities $p_\sigma(x)$ for motion in
the forward direction after a particle appears in state $\sigma$ at
position $x$. Similarly, $1-p_\sigma(x)$ is the probability for
backward motion, see Fig. \ref{f:prob}. Since the two potentials are
shifted with respect to each other $p_2(x)=p_1(x-\delta)$. The
probability $p_+$ requires two subsequent forward movements of this
type, $p_-$ results from two backward movements. As described in
appendix \ref{a:q}, the probabilities $p_\sigma(x)$ can be calculated
approximatively for large $U/T$. Fig. \ref{f:eff_nl} (a) shows the
efficiency estimated by Eqns. (\ref{eq:etaapp}) and (\ref{eq:ppm})
together with the numerically obtained values for comparison. For
small forces the agreement is good, thus confirming our simplified
picture of energy transduction in this regime.

\section{Concluding remarks}

In the previous sections, we have studied the efficiency of energy
transduction from chemical energy to mechanical work using a simple
two-state model under isothermal conditions.  We considered three
different examples: system A with two shifted potentials and both
active and thermal transitions between the two states localized at the
potential minima; system B with an additional symmetry between the
two states and no passive thermal transitions; and finally system C
with a flat weakly bound state and non-localized passive transitions.
We demonstrated that energy transduction can be very efficient in the
systems A and B with localized transitions and shifted potentials
and is at least two orders of magnitude smaller in system C which
requires diffusive steps for motion to occur.  Interestingly, the
largest efficiency can occur far from equilibrium.  This is in
particular the case for systems A and C which both are not very
efficient in the linear response regime.

\subsection{Isothermal motors, heat engines and Brownian ratchets}

Efficiencies of energy transduction have been studied and discussed
for a long time. Of particular significance is the concept of Carnot
which defines the efficiency of macroscopic heat engines coupled to
two thermal baths at temperatures $T^\pm$ and $T^+>T^-$ as
\begin{equation}
\eta_{\rm Carnot}=-\frac{\fe v}{\dot Q^+} \quad , \label{eq:etaC}
\end{equation}
where $\dot Q^+$ is the rate of heat transfer from the {\em hot}
reservoir. This definition then leads to an upper limit of the
efficiency $\eta_{\rm Carnot} \leq (T^+-T^-)/T^+$ which cannot be surpassed by
any heat engine. In order to characterize energy transduction in
biological systems, a natural choice is \cite{hill74}
\begin{equation}
\eta=-\frac{\fe v}{r \Delta\mu} \quad ,
\end{equation}
which we have adopted in this paper, see Eq. (\ref{eq:eff}), and which
is based on the chemical potential difference between fuel and
reaction products. As we have discussed, this efficiency obeys $\eta
\leq 1$ in order to satisfy the first law of thermodynamics, but there
is no nontrivial upper bound.  In addition to the obvious fact that
$\eta_{\rm Carnot}$ describes a heat engine and $\eta$ an isothermal motor, there
remains a fundamental difference between the two cases: the definition
of $\eta_{\rm Carnot}$ assumes that all heat dissipated in the bath $T^-$ is
lost.  This is true in most practical cases, however if the bath at
$T^-$ was also used as the hot bath of a second heat engine, some of
this energy could in principle be reused. Similarly, the
definition of $\eta$ takes into account the energy of the lower-energy
reservoir, thus assuming that the energy of the reaction products
remains available. One might think that it is possible to avoid this 
difference between the two definitions by choosing: 
\begin{equation}
\eta' = -\frac{\fe v}{r \mua}
\end{equation}
where $\mua$ would be the chemical potential of the fuel (ATP).  This
definition would share with Carnot's definition the viewpoint that the
energy of the reaction products are not useful, and since $\eta'=\eta\Delta\mu/\mua$ would lead to the upper bound $\eta'\leq
(\mua-\mup)/\mua$. Such a choice, however, suffers from a serious
shortcoming: only chemical potential differences are physically
meaningful. Depending on the state of reference used for defining
$\mua$, the value of $\mua$ could be positive, negative or even
vanish.  

The example given above demonstrates that comparing efficiencies can
be dangerous as they may be based on different definitions
corresponding to different points of view. This is also the case for
ratchet models which have been studied in many variants and under
widely varying physical conditions.  All definitions described above
have been used in the literature: The definition $\eta_{\rm Carnot}$
for systems driven by temperature differences
\cite{feyn66,seki97,soko97,mats98,dere98}, the definition for $\eta$ given by
Eq. (\ref{eq:eff}) \cite{hill74,juli97B,juli98} as well as $\eta'$
\cite{shib97}.  Alternative definitions 
have been proposed for situations where the chemical reaction is not
fully specified \cite{juli95,parr98}. Other definitions of energy
transduction efficiencies have been used for systems which are driven
by stochastic or deterministic forces
\cite{seki97,hond98,kame98,seki98}. Recently, 
Sekimoto has presented a unified picture which includes most systems
in a common framework \cite{seki97}.  However, in general, a given
definition is adapted to one particular physical situation.

\subsection{The two-state model and biological motors}

One important motivation of this work is to clarify the general
properties of energy transduction of biological motors.  The
characteristic behaviors of our system A and B with localized
transitions and shifted potentials are similar to those observed for
processive biological motors such as e.g. kinesins which move along
microtubules and for which the consumption of ATP and the subsequent
stepping are strongly correlated for small external forces
\cite{svob93,schn97,hua97}.  Kinesin motors consist of two identical
active head groups which both hydrolize ATP \cite{mand99}.  There is
evidence suggesting that the motor could ``walk'' in a head-over-head
fashion along microtubules, detaching a head in the back and
reattaching in front of the molecule while keeping the second head
bound \cite{bloc98,hanc98}. In such a picture each ATP-hydrolysis
cycle leads to a new situation where both heads have exchanged their
roles and the center of mass of the molecule has advanced one filament
period. This type of motion is captured in a simple way in the variant
B of our model which is symmetric with respect to the two
states. Because of this symmetry, both states are indistinguishable
but the corresponding potentials are shifted by $l/2$:
$W_2(x)=W_1(x-l/2)$. We therefore identify each of the two states with
one kinesin head and $l/2$ with the filament period, see
Fig. \ref{f:pot} (b). Recently, system C with a structureless excited
state has been used for single kinesin heads which were observed to
move processively along a microtubule \cite{okad99}.  Models of this
type have typically been considered in the context of non-processive
motors such as myosins which have a weakly bound state during their
interaction cycle.  Myosins interact with a filament to generate
displacements of the order of several nm, but they do not continuously
move along a filament as individual motors since they easily lose
their track and diffuse away \cite{spud90,howa97}.  The latter
phenomenon is not captured in the one-dimensional two-state model,
however the flat potential of system C requires diffusive steps for
average motion and the efficiency is therefore smaller than for
systems A and B.  Under physiological conditions non-processive motors
operate not as isolated enzymes but move together in large groups. In
this situation, however, diffusive steps become unimportant and the
efficiency becomes large and reaches the same orders of magnitude as
for model A and B described here \cite{davi69,hill80,juli95}.

\begin{figure}
\centerline{\psfig{file=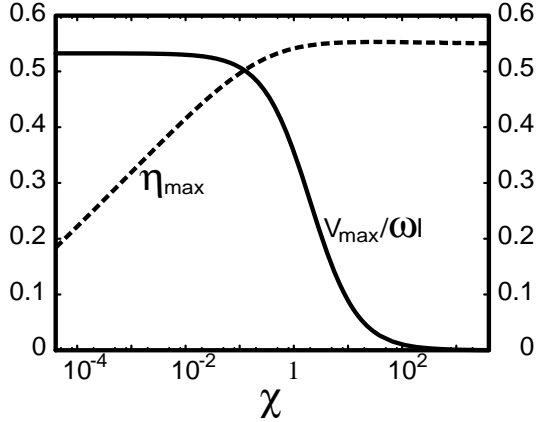,width=7.0 cm}}
\vspace{0.5cm}
\caption{
Maximal efficiencies (broken line) and normalized velocities (solid
line) as a function of $\chi$ for system B and $\Delta\mu/U=0.6$ and
$T/U=0.05$.}
\label{f:eff_vmax}
\end{figure}

When comparing our simple models with biological motors, the value of
the adimensional parameter $\chi=\trans\frii l^2/U$ introduced in
Eq. (\ref{eq:chi}) is crucial. The relevant orders of magnitude for
most parameters are well known: Energy scales are $U\simeq 10 T$
\cite{hill74}, typical time scales of conformational changes are
$\trans^{-1}\simeq 1$ms and the relevant length scale is $l\simeq
5-10$nm \cite{spud90}. However, the friction coefficient $\frii$ is
unknown and difficult to estimate.  Therefore, we do not know at which
value of $\chi$ biological motors operate. The role of $\chi$ on the
functioning of the system can be discussed by comparing both the
maximal efficiency and the dimensionless velocity $v/\trans l$ as a
function of $\chi$, see Fig. \ref{f:eff_vmax}. 
The diagram reveals that for large values of $\chi$ for which the
efficiency is large the velocities become small. For small $\chi$
velocities are optimal but efficiency becomes negligible.  This
observation suggests that optimal conditions are obtained in the
intermediate regime $\chi\simeq 0.1-1$ where chemical times and
sliding times along the potential slopes become comparable.
If linear molecular motors operate in this regime, the microscopic 
friction coefficient $\frii$ is of the order of $10^{-7}-10^{-6}$ kg/s.
If we estimate $\frii$ from simple hydrodynamic arguments 
($\frii_{\rm h} \simeq 6 \pi \eta_{\rm vis} l$), where $\eta_{\rm
vis}$ is some measure of a ``local'' viscosity, we find $\eta_{\rm
vis}\simeq 10 - 100$ Poise, $10^3 - 10^4$ times the viscosity of 
water, values compatible with dense macromolecular solutions.
Interestingly, this order of magnitude, corresponds to a diffusion
coefficient of $4 \cdot 10^{-14}$ m$^2$/s, a value reported recently for 
single headed kinesin \cite{okad99}.
This observation together with our estimate suggest that linear molecular 
motors are optimized both from the velocity and the efficiency standpoint.  

\acknowledgements

We acknowledge stimulating discussions with S. Camalet, R. Everaers,
P.G. de Gennes, K. Sekimoto and T. Shibata.

\appendix

\section{Onsager coefficients}
\label{s:ons_rel}
In linear response theory, the behavior of the system is completely
defined by the Onsager coefficients $\lambda_{ij}$.  The Onsager
symmetry relation (\ref{eq:ons_rel}) follows from general
thermodynamic arguments and the microscopic reversibility. The
calculation of Onsager coefficients is difficult, however the symmetry
relation can be verified by general arguments as shown in section
\ref{s:ons_rel_sp}. In section
\ref{s:ons_rel_mp} we obtain explicit expressions for the coefficients
$\lambda_{ij}$ for a many-motor system as introduced in
\cite{juli97B,juli95}.

\subsection{Symmetry relation for a single motor}
\label{s:ons_rel_sp}
In order to demonstrate the symmetry relation of Onsager coefficients
for the two-state model, we start from the probability distributions
at equilibrium ($\fe=0$, $\Delta\mu=0$) as
\begin{equation}
\label{eq:peq}
P_\sigma^{eq} = N e^{-W_\sigma(x)/T} \quad ,
\end{equation}
with a normalization factor
\begin{equation}
N^{-1} \equiv \sum_\sigma \int_0^l dx\;  e^{-W_\sigma/T}
\quad .
\end{equation}
For small $\fe l/T\ll 1$ and $\Delta\mu/T\ll 1$
we define deviations $p_\sigma(x)$ from equilibrium
which obey
\begin{equation}
P_\sigma(x) = N e^{-W_\sigma/T}\left( 1 + p_\sigma(x)\right
)\quad .
\end{equation}

Without loss of generality we consider the case where only
the transitions $\alpha_\sigma$ and $\beta_\sigma$ occur but
$\gamma_\sigma=0$.
To linear order in $\Delta\mu$, the transition rates can be written
as
\begin{eqnarray}
\omega_1(x)& = &\left (\omega(x) + \frac{\alpha(x)\Delta\mu}{T}\right
)
e^{W_1/T} \nonumber \\
\omega_2(x) & = &\omega(x)e^{W_2/T}
\end{eqnarray}
with $\alpha=\alpha_2 e^{-W_2/T}$ and $\omega(x)\equiv (\alpha_2 +
\beta_2) e^{-W_2/T}$.  Using Eq. (\ref{eq:Psp}) we
find to linear order
\begin{eqnarray}
-\frac{ T}{\frii} 
\partial_x \left(e^{\frac{-W_1(x)}{T}} \partial_x p_1 \right )
+ \omega(x) (p_1-p_2) & = & h_1(x)\nonumber \\
-\frac{ T}{\frii} 
\partial_x \left(e^{\frac{-W_2(x)}{T}} \partial_x p_2 \right )
- \omega(x) (p_1-p_2) & = & h_2(x) \quad .\label{eq:linph}
\end{eqnarray}
The fields $h_\sigma(x)$ are nonzero in the
presence of mechanical or chemical forces:
\begin{eqnarray}
h_1(x)& =&  - \alpha(x) \Delta\mu/T
- \partial_x e^{-\frac{W_1(x)}{T}}  \fe/\frii\nonumber\\
h_2(x)& =&\alpha(x)\Delta\mu/T
- \partial_x e^{-\frac{W_2(x)}{T}}  \fe  /\frii
\quad . \label{eq:ph}
\end{eqnarray}
Eq. (\ref{eq:linph}) represents a linear relation 
between $p_\sigma$ and $h_\sigma$ which can be inverted
and which thus defines a response kernel
\begin{equation}
p_\sigma(x) = \sum_\rho \int_0^l dx' \chi_{\sigma\rho}(x,x') h_\rho(x')
\quad .
\label{eq:linhp}
\end{equation}
This allows us to express the velocity and the fuel consumption rate
within linear response theory:
\begin{eqnarray}
v&=&\int_0^l dx\; \left[ \fri e^{-\frac{W_1}{T}}\left(
 \fe -T \partial_x p_1\right) \right. + \nonumber \\
& & +  \left. \fri e^{-\frac{W_2}{T}}\left(
 \fe - T \partial_x p_2 \right ) \right] \label{eq:vsp}\\
r&=&\int_0^ldx\; \alpha(x) \left[(p_1-p_2) + \Delta\mu/T \right]
\label{eq:rsp}
\end{eqnarray}
The Onsager coefficients $\lambda_{ij}$ can be written in terms of the
response functions $\chi_{\sigma\rho}(x,x')$.  In particular, we find
for the coefficients of mechano-chemical coupling
\begin{equation}
\begin{array}{lll}
\lambda_{12} & \equiv & {\displaystyle \frac{\partial v}{\partial\Delta\mu}} \nonumber \\
& = &
{\displaystyle \!\!\!\!\! \int_0^l \!\! dx \!\! \int_0^l dx'\!\! }
 \left [  \fri e^{-\frac{W_1(x)}{T}} (\partial_x \chi_{11}(x,x')
- \partial_x \chi_{12}(x,x')) \right .\nonumber \\
& + & \left.  \fri e^{-\frac{W_2(x)}{T}} (\partial_x \chi_{21}(x,x')
- \partial_x \chi_{22}(x,x'))\right ] \alpha(x')\label{eq:l12} 
\end{array}
\end{equation}
\begin{equation}
\begin{array}{lll}
\lambda_{21} & \equiv &{\displaystyle \frac{\partial r}{\partial \fe }} \nonumber \\
 & = &{\displaystyle \!\!\! - \!\!  \int_0^l \!\! dx \!\!
\int_0^l dx'}
 \left [ (\chi_{11}(x,x')
- \chi_{21}(x,x')) \fri \partial_{x'}e^{-\frac{W_1(x')}{T}}  \right
.\nonumber \\
& + & \left.   (\chi_{12}(x,x')
- \chi_{22}(x,x'))\fri \partial_{x'}e^{-\frac{W_2(x')}{T}}\right ] \alpha(x).\label{eq:l21}
\end{array}
\end{equation}
Performing a partial integration in Eq. (\ref{eq:l12}), we find that
the Onsager symmetry relation (\ref{eq:ons_rel}) is satisfied exactly
if the response functions obey the symmetry relation
\begin{equation}
\chi_{\sigma\rho}(x,x')=\chi_{\rho\sigma}(x',x) \quad .
\label{eq:symchi}
\end{equation}

This symmetry relation follows from the hermiticity of the linear
operator ${\cal L}$ defined in Eq. (\ref{eq:linph}) which can be expressed
as
\begin{equation}
{\cal L} {p_1 \choose p_2} = {h_1 \choose h_2} \quad ,
\end{equation}
where
\begin{equation}
{\cal L} = \left (
\begin{array}{c c}
{\cal L}_1 +\omega & -\omega \\
-\omega & {\cal L}_2 + \omega 
\end{array}
\right )
\label{eq:lm}
\end{equation}
with
\begin{equation}
{\cal L}_i = -\fri T \partial_x e^{\frac{-W_i(x)}{T}} \partial_x \quad .
\end{equation}
The operator ${\cal L}$ is hermitian since the matrix (\ref{eq:lm}) is
symmetric and ${\cal L}_i$ itself is hermitian.  The latter is easily
verified by partial integration:
\begin{equation}
\!\!\!\!\!\!\int_0^l\! dx q(x) (\partial_x e^{\frac{-W_1(x)}{T}} \partial_x p(x))
=\! \int_0^l\! dx p(x) (\partial_x e^{\frac{-W_1(x)}{T}} \partial_x q(x))  .
\end{equation}

\subsection{Onsager coefficients for
many rigidly coupled motors}
\label{s:ons_rel_mp}

Onsager coefficients can be calculated explicitly for a model of
rigidly coupled motors. In this model, the two states are defined in
the same way as before, but the current term in Eq. (\ref{eq:Psp}) is
replaced by a convective term since all particles move with the same
velocity.  The transport equations are given by \cite{juli95,juli97B}
\begin{eqnarray}
\partial_t P_1 + v \partial_{x} P_1& =& -
\omega_1(x) P_1 + \omega_2(x) P_2 \nonumber\\
\partial_t P_2 + v \partial_{x} P_2& =& 
\omega_1(x) P_1 - \omega_2(x) P_2 \quad , \label{eq:Pspmp}
\end{eqnarray}
the velocity $v$ is determined
by the force balance condition
\begin{equation}
v=\fri\left [\fe -\int_0^l dx\;(P_1 \partial_x W_1+
P_2\partial_x W_2) \right ]\quad . \label{eq:vmp}
\end{equation}
Steady state distributions $P_1$ and
$P_2=1/l-P_1$ are solutions to
\begin{equation}
v\partial_x P_1 = (\omega_1+\omega_2) P_1 + \omega_2/l \quad
.\label{eq:deq}
\end{equation}
Using a power expansion in the velocity, $P_1$ can be written
to lowest order
\begin{equation}
P_1(x)=P_1^{(0)}(x)+P_1^{(1)}(x)  v + O(v^2) \quad,
\end{equation}
with $P_1^{(0)}=\omega_2/(\omega_1+\omega_2)l$, and
\begin{equation}
P_1^{(1)}=-\frac{1}{\omega_1+\omega_2}\partial_x P_1^{(0)} \quad
.\label{eq:P11}
\end{equation}
As in the last section we use a reaction scheme with $\gamma_\sigma=0$
in order to keep the expressions simple. Also, without loss of
generality we consider the case where $\alpha_1$ depends on
$\Delta\mu$ but $\alpha_2$ remains constant. For small $\Delta\mu/T\ll
1$, we express the transition rates of Eq. (\ref{eq:om1om2}) as
\begin{equation}
\omega_1 = \omega_2 e^{-\Delta W / T}
(1+\bar{\alpha}) \quad ,
\end{equation}
where $\bar{\alpha}\equiv \alpha_2\Delta\mu/\omega_2 T$.
The force-velocity relation for small $v$ is given by
\begin{equation}
\fe=f^{(0)} + (\fri+f^{(1)})v + O(v^2) \quad ,
\label{eq:fexp}
\end{equation}
with coefficients
\begin{equation}
f^{(n)}=-\int_0^l dx P_1^{(n)} \partial \DW \quad .\label{eq:fn}
\end{equation}
which depend on $\Delta\mu$. We can now calculate the Onsager
coefficients. The effective friction 
\begin{equation}
\lambda_{11}\equiv\left. \frac{\partial v}{\partial \fe}\right|_{\Delta\mu=0} 
\label{eq:l11}\quad,
\end{equation}
can be determined from Eq. (\ref{eq:fexp}):
\begin{equation}
\lambda_{11}^{-1}=\fri+
\frac{1}{lT}\int_0^l dx\frac{e^{-\DW/T}(\partial_x\DW)^2}{\omega_2
(1+e^{-\DW/T})^3} \quad .
\end{equation}
Similarly,
\begin{equation}
\lambda_{12}\equiv\left.\frac{\partial v}{\partial \Delta\mu}\right|_{\fe=0}
=-\lambda_{11} \left.\frac{\partial \fe}{\partial \Delta\mu}\right |_{v=0}
\quad ,\label{eq:lc12}
\end{equation}
leads to
\begin{equation}
\lambda_{12}=-\lambda_{11} \left.\frac{\partial
f^{(0)}}{\partial\Delta\mu}
\right |_{\Delta \mu=0} = -\frac{\lambda_{11}}{l T}\int_0^l dx
\frac{\alpha_2 e^{-\DW/T} \partial_x\DW}{\omega_2(1+e^{-\DW/T})^2} 
\end{equation}
The second cross-coefficient
\begin{equation}
\lambda_{21}\equiv\left. \frac{\partial r}{\partial \fe}\right|_{\Delta\mu=0}
= \lambda_{11}\left. \frac{\partial r}{\partial v}\right |_{\Delta\mu=0}
\quad , \label{eq:lc21}
\end{equation}
is determined from the fuel consumption rate $r(v,\Delta\mu)$.  Using
Eqns. (\ref{eq:r}), (\ref{eq:lc21}) and (\ref{eq:P11}), we obtain
\begin{eqnarray}
\left. \frac{\partial r}{\partial v}\right|_{\Delta\mu=0} & 
 = &  \left. \int_0^l dx \left (\alpha_2 +\alpha_1) \partial_v P_{1} \right] 
\right |_{\Delta \mu=0} \nonumber \\
& = &  - \frac{1}{lT}  \int_0^l dx \frac{\alpha_2 \partial_x(\DW) e^{-\DW/T}}
{\omega_2(1+e^{-\DW/T})^2}
\quad , \label{eq:drdv}
\end{eqnarray}
and thus as required $\lambda_{12}=\lambda_{21}$.  Finally,
\begin{eqnarray}
\lambda_{22} & =& \left. \frac{\partial r}{\partial \Delta\mu}\right
|_{\fe=0}
= \lambda_{12}\left. \frac{\partial r}{\partial v}\right |_{\Delta \mu=0}
+ \left . \frac{\partial r}{\partial \Delta\mu}\right |_{v=0} \nonumber \\
 & = & \frac{\lambda_{12}^2}{\lambda_{11}} 
+ \left . \frac{\partial r}{\partial \Delta\mu}\right |_{v=0} \quad ,
\label{eq:l22}
\end{eqnarray}
which requires to calculate
\begin{equation}
\!\!\! \left. \frac{\partial r}{\partial \Delta\mu}\right |_{v=0} 
= \int_0^l\left[ \frac{\partial P_1^{(0)}}{\partial \Delta\mu}
(\alpha_1+\alpha_2)+P_1^{(0)}\frac{\partial}{\partial\Delta\mu}(
\alpha_1+\alpha_2)\right].
\end{equation}
Using Eqns (\ref{eq:om1om2}) and (\ref{eq:r}), we obtain
\begin{equation}
\left. \frac{\partial r}{\partial \Delta\mu}\right |_{v=0}
= \frac{1}{lT}  \int_0^l dx \left[\frac{\alpha_2 e^{-\DW/T}}{(1+e^{-\DW/T})}\left(1-\frac{\alpha_2}{\omega_2}\right)\right] \quad .
\label{eq:rmv}
\end{equation}
Note that both $\lambda_{11}$ and $\lambda_{22}$ are positive while
$\lambda_{12}$ can have either sign.

\section{Diffusion close to a potential maximum}
\label{a:q}

In Section \ref{s:smallf}, we introduced the probabilities
$p_\sigma(x)$ that a particle which initially starts at position $x$
close to a potential maximum will finally escape in the positive
direction. We will calculate this probability for a piecewise linear
potential as shown in Fig. $10$ in the limit where
the potential slopes extend to infinity which corresponds to large
potential amplitudes $U/T\gg 1$. 
We consider the Fokker-Planck Equation
\begin{equation}
\partial_t P + \partial_x J=0 \quad ,
\end{equation}
with 
\begin{equation}
J=-\fri [T\partial_x P +
 P \partial_x W -P \fe] \quad ,\label{eq:current}
\end{equation}
for initial condition
\begin{equation}
P(x,t=0)=\delta(x-x_0) \quad .
\end{equation}
In order to determine what fraction of 
particles move to the right after a long time, we define
the Laplace transform 
\begin{equation}
\tilde P(x,s)=\int_0^{\infty} dt P(x,t) e^{-st} \quad 
\end{equation}
of the distribution and $\tilde J(x,s)=-\fri[T\partial_x \tilde P
+\tilde P\partial_x W-\tilde P\fe]$ of the current.
\begin{figure}
\centerline{\psfig{file=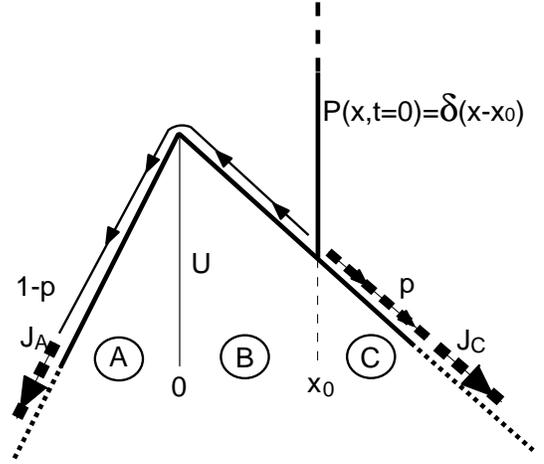,width=7.0 cm}}
\vspace{0.5cm}
\caption{Schematic representation of diffusion near a potential
maximum which can be divided in three different regions A, B and C.
For a particle which initially appears at $x=x_0$, we are interested
in the probability $p$ that it will finally move forward. This
probability is related to the current $J_C$ which can be calculated in
the limit of long times.}
\label{f:configuration}
\end{figure}
The average number of particles which pass at position $x$ after long times
is given by
\begin{equation}
\int_0^\infty dt J(x,t) = \tilde J(x,s=0) \quad .
\end{equation}
Noting that 
\begin{equation}
\int_0^\infty dt e^{-st} \frac{\partial}{\partial t} P(x,t)=
-P(x,0) + s \tilde P(x,s) \quad ,
\end{equation}
we obtain an equation for $\tilde P$:
\begin{equation}
\label{eq:lap2}
\fri \partial_x \left [T \partial_x \tilde{P} + (\partial_x W -\fe)
 \tilde{P}\right ]-s \tilde{P}=-P(x,0) \quad . \label{eq:lap21}
\end{equation}
Since we are interested in $s=0$ we have to solve
\begin{equation}
 \partial_x \left [T \partial_x \tilde{P} + (\partial_x W -\fe)
 \tilde{P}\right ]=-\frii \delta(x-x_0) \quad .\label{eq:lap20}
\end{equation}
All quantities of interest can be easily calculated if we we assume a
piecewise linear potential
\begin{equation}
W(x) = \left\{
\begin{array}{l l}
U + \fm x & \quad x<0 \\
U - \fp x & \quad x \geq 0 
\end{array}
\right .\quad  ,
\end{equation}
with the potential slopes $\fm=U/a$ and $\fp=U/(l-a)$.  We
distinguish three different regions A, B and C along the $x$-axis, see
Fig. $10$. Within each region, the solution to
Eq. (\ref{eq:lap20}) is
\begin{equation}
\label{eq:lap3}
\tilde{P}(x)= C_0 + C_1 e^{-(W(x)-x\fe)/T} \quad ,
\end{equation}
with two constants $C_0$ and $C_1$ which have to be determined for
each of the three regions. We denote the corresponding solutions
$\tilde P_A$, $\tilde P_B$ and $\tilde P_C$.  Since we are looking for
solutions which do not diverge for $x\rightarrow \pm\infty$, we have
$C_1^A=C_1^C=0$ in regions A and C and therefore $\tilde P_A=C_0^A$
and $\tilde P_C = C_0^C$.  This boundary condition for large $x$ can
also be derived more carefully by first imposing the condition
\begin{equation}
\lim_{x \rightarrow\pm\infty} \tilde P(x,s) =0
\end{equation}
for $s>0$ and taking the limit $s\rightarrow 0$ afterwards.
Additional boundary conditions are the conditions of continuity
of $\tilde P(x,0)$ at $x=0$ and $x=x_0$
\begin{eqnarray}
\tilde P_A(0)&=&\tilde P_B(0) \nonumber \\
\tilde P_B(x_0)&=& \tilde P_C(x_0) \quad ,
\end{eqnarray}
and the matching conditions
\begin{eqnarray}
\partial_x \tilde P_A(0)& =& \partial_x\tilde P_B(0) -\tilde P_A(0)(\fp+\fm)/T
\nonumber \\
\partial_x \tilde P_B(x_0) & = & \partial_x \tilde P_C(x_0) +\frii/T \quad ,
\end{eqnarray}
which follow from integrating Eq. (\ref{eq:lap20}) at the
singularities of the potential slope at $x=0$ and the delta-function 
at $x=x_0$. With these conditions, all free parameters
can be determined and we obtain
\begin{eqnarray}
 C_0^A & = & \frac{\frii}{\fm+\fp}e^{-(\fp +\fe)x_0/T} \\
 C_0^C & = & \frac{\frii}{\fp+\fe}
\left[ 1-\frac{\fm-\fe}{\fm+\fp}e^{-(\fp +\fe)x_0/T}\right] ,
\end{eqnarray}

The corresponding currents $\tilde J(x,s=0)$ are constant in region
A and C:
\begin{eqnarray}
 \tilde{J}_A & = & -\fri (\fm-\fe)C_0^A \\
 \tilde{J}_C & = & \fri (\fp+\fe)C_0^C \quad ,
\end{eqnarray}
which satisfy the normalization condition $J_C-J_A=1$.
The probability $p(x_0)$ for forward motion of a particle which initially was
at $x_0$ is equal to $J_C$:
\begin{equation}
\label{eq:ptot}
 p(x_0) = 1-\frac{\fm-\fe}{\fm + \fp}e^{-(\fp +\fe)x_0/T} 
\label{eq:psigma}\quad .
\end{equation}
The probabilities introduced
in Eq. (\ref{eq:ppm}) are given by $p_2(0)=p(l-\delta)$, $p_1(\delta)=
p(\delta-a)$.


\begin{thebibliography}{99}
\bibitem{footnote1} The quantities $v$ and $r$ are either ensemble averages or time averages. Experiments on single molecules and finite observation time require a special analysis (see for example \cite{seki98}).
\bibitem{footnote2} We assume that $\Delta\mu$ is changed by varying the
ATP concentration while keeping $\mu_{ADP}$ and $\mu_P$ fixed.
\end{thebibliography}
\end{document}